\documentclass[aps, amsmath,amssymb,prl,twocolumn,showpacs]{revtex4-1}
\usepackage{graphicx,color,dcolumn}
 \usepackage[utf8]{inputenc}
\usepackage[T1]{fontenc}
\usepackage{siunitx}
\usepackage{hyperref}
\usepackage{ulem}

\begin{document}

\title{Non-linear fate of internal wave attractors}

\author{Hélène Scolan$^1$, Eugeny Ermanyuk$^{1,2}$  and Thierry Dauxois$^1$
}
\affiliation{
1. Laboratoire de Physique de l'\'Ecole Normale Supérieure de Lyon, Université de Lyon, CNRS, 46 Allée d'Italie, F-69364 Lyon cedex 07, France.\\
2. Lavrentyev Institute of Hydrodynamics, Novosibirsk State University, Novosibirsk, Russia}


\date{\today}

\begin{abstract}
We present a laboratory study on the instability of internal wave attractors in a trapezoidal fluid domain filled with uniformly stratified fluid.
Energy is injected into the system via standing-wave-type motion of a vertical wall.
Attractors are found to be destroyed by parametric subharmonic instability (PSI) 
via a triadic resonance which is shown to provide a very efficient energy pathway
from long to short length scales. This study provides an explanation why attractors 
may be difficult or impossible to observe in natural systems subject to large amplitude forcing.
\end{abstract}


\maketitle



{\bf Introduction.} Energy transfer from large to small-scale is a critical issue in the
dynamics of large geophysical systems such as ocean and atmosphere.
In this context, internal waves are of particular interest due to their specific dispersion and
reflection properties. In a uniformly stratified fluid of infinite extent,
which is the usual simplification of a realistic slow-varying stratification, internal waves
propagate as oblique beams obeying~\cite{Phillips1966} the following dispersion
relation
$\theta=\mbox{arccos}{(\omega/N)}$, where $\theta$ is the angle between the wave beams and
the vertical, $\omega$ the wave frequency, $N=[-(g/\rho)(d\rho/dz)]^{1/2}$ the constant buoyancy frequency,
and $\rho(z)$ the density stratification a function of the vertical coordinate~$z$.
Consequently, the beam angle with respect to the vertical is preserved when the beam is reflected at
a rigid boundary. These restrictive conditions give a purely geometrical reason for strong variations
of scale (focusing or defocusing) when an internal-wave beam is reflected at a sloping boundary.
The complex dynamics of this phenomenon has been extensively studied~\cite{Dauxois1999,Swinney2008}.

\begin{figure}

\null\hskip -0.475truecm\includegraphics[width=4.25cm]{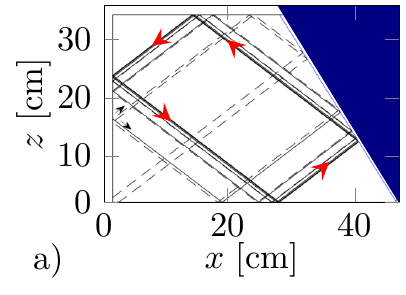}
\hskip-0.15truecm
\includegraphics[width=4.75cm]{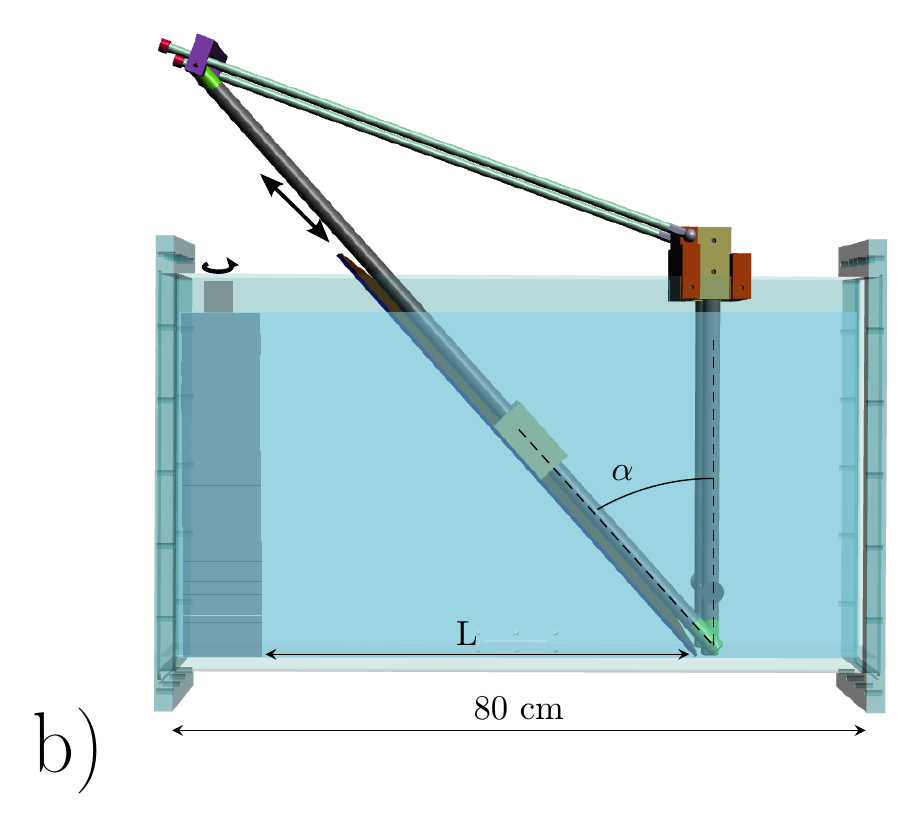}

\caption{(Color online)
Panel (a) presents the prediction of a wave-ray billiard with a bottom length 
$L=\SI{45.6}{cm}$ and a depth $H=\SI{32.6}{cm}$,
while the sloping wall is inclined at an angle $\alpha=30^{\circ}$ with the vertical. The ray direction (defined by the group velocity) of the limit-cycle is counter-clockwise as shown by red arrows.
Panel (b): Corresponding experimental setup showing the wave generator and the sloping wall inside the immobile tank of size $80\times17\times42.5$ cm.
The working bottom length of the section, the depth and
 the sloping angle are the one given in panel (a).
Conventional double-bucket technique is used to create a uniform
stratification with a buoyancy frequency $N=\SI{0.95}{rad/s}$.
\label{Experimentasetup}}
\end{figure}

In confined fluid domains, focusing usually prevails: successive reflections of internal
wave beams at rigid boundaries produce interestingly nearly closed loops which gradually
converge toward a closed trajectory, an internal wave attractor~\cite{Maas1995}.
Ray trajectories in arbitrary shaped containers are generally not closed, and therefore energy injected in the domain is evenly distributed. On the contrary, when an attractor is present as in Fig.~\ref{Experimentasetup}a, essentially all the energy is concentrated on a few beams defining the limit cycle and, consequently, injected
energy being focused, nonlinear instabilites are more likely to be expected.
Experimentally, an attractor was first demonstrated in a trapezoidal domain
filled with uniformly stratified fluid~\cite{Maas1997}.
Simplistic considerations of wave-ray billiard lead to the unphysical conclusion of
vanishingly small width of attractor branches (infinite focusing). In reality, a finite
width of wave beams is set by the balance between geometric focusing and viscous
broadening~\cite{Hazewinkel2008,Grisouard2008}.
Attractors were shown to be sufficiently
robust to be observable in a non-uniform stratification and in test tanks with corrugated
walls~\cite{Hazewinkel2010} as well as in laterally infinite fluid domains with appropriate
bottom topography~\cite{Echeverry2011}.
The significance of wave attractors has been recognized in rotating
fluids~\cite{Rieutord2000} and proposed for magnetized materials~\cite{Maas2005}.

Theoretical studies on the behavior of a hyperbolic system describing attractor-like
structures in confined domains reveal highly complicated dynamics. However,
this rich dynamics arises in strictly linear PDEs, which form the background of existing
theoretical studies~\cite{Maas2005}. Numerically, nearly all studies of wave
attractors solve linear equations of motion as stressed in~\cite{Grisouard2008}.
Experimentally, attractors are usually generated by low-amplitude vertical or horizontal
oscillations of test tanks filled with stratified fluids~\cite{Maas1997,Hazewinkel2008,Hazewinkel2010}
or by a modulation of the angular velocity in rotating fluids~\cite{Manders2003}.
Oscillations of small objects have also been used to produce internal waves forming attractor-like
patterns in 2D~\cite{Maas2005} and 3D~\cite{Hazewinkel2011} geometries.
Experimentally observed attractors had therefore relatively low energy and their behavior can be
explained by linear mechanisms. In this connection, a number
of important questions arise. What happens to wave attractors as the amount of injected
energy increases? What is the main mechanism of instability which destroys
wave attractors? Does the instability produce new length scales which are shorter than
the equilibrium width~\cite{Hazewinkel2008} of attractor? What is beyond the instability?
In the present letter, we address all these issues experimentally.

\begin{figure*}[htb]
\includegraphics[width=\linewidth]{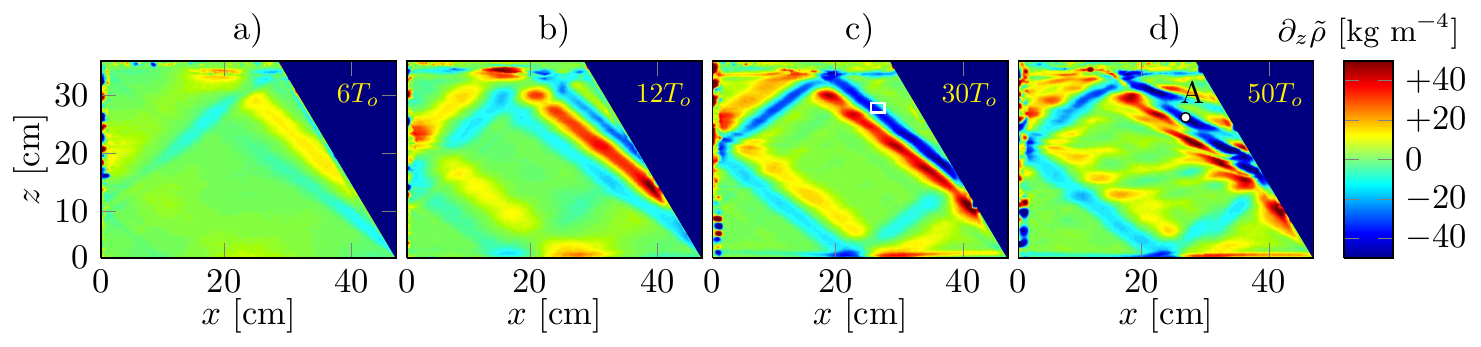}
\vskip -0.4truecm\caption{(Color online) Snapshots of the vertical density gradient field  for $t=6T_0$~(a),  $12T_0$~(b),  $30T_0$~(c),
$50T_0$~(d) where $T_0=2\pi/\omega_0$ is the primary wave period.   Note that the shade (color online) scale is
the same in all panels. The wave frequency is  $\omega_0/N=0.62 \pm 0.02$
and the motion amplitude of the plates of the generator is set to $a=0.25$~cm. The small white rectangle in panel (c) defines the acquisition region used for computing the time-frequency spectrum used for Fig.~\ref{figtimefrequency}.}
\label{rawfield}
\end{figure*}

{\bf Experiment.} To generate internal wave attractors with a high level of injected energy, we use a novel approach,
 presented in Fig.~\ref{Experimentasetup}b.
Experiments are performed in a quiescent test tank.
The classic trapezoidal geometry~\cite{Maas1997} is designed with a sliding
sloping wall which can be slowly inserted into the fluid once the test tank is filled.
The energy is injected into the experimental system by the internal wave 
generator~\cite{Gostiaux2007,Mercier2010,Joubaud2012}
 tuned to produce the first vertical mode for internal waves in finite depth~$H$.
The time dependent profile of the generator, and therefore of the left side of the tank,
is given by
\begin{equation}
\eta(z,t)=a\cos(\pi z/H)\cos(\omega_0 t)\,,\label{profile}
\end{equation}
where $a$ is its amplitude. The profile~(\ref{profile}) is reproduced in discrete stepwise
form by the motion of 51 horizontal plates driven by the rotation of a vertical camshaft. 
Since the thickness of each plate is small compared to the width of the wave-attractor 
beams, the discretization does not produce any secondary perturbations to the
wave field, in agreement with~\cite{Hazewinkel2010,Mercier2010}. The perturbations 
of the density gradient are evaluated with the synthetic schlieren
technique~\cite{Dalziel2000} from apparent displacements of elements of the background
random dot pattern placed behind the test tank.
A series of experiments has been performed varying the parameters $a\in[0.15,0.5]$ cm, 
$\omega_0/N\in[0.49,0.83]$ and $\alpha=15$ or 30$^\circ$.
We will emphasize now the cases  $\alpha=30^\circ$ and $\omega/N=0.62$,
but it is important to stress that all results are fully reproducible and lead to similar conclusions.

{\bf Results and discussion.} 
The evolution of observed internal wave patterns with time
is presented in {Fig}.~\ref{rawfield} for a moderately large amplitude~$a=0.25$ cm.
One can see that the attractor reaches its fully developed state after a transient of 
roughly 30 periods of oscillation of the generator $T_0=2\pi/\omega_0$. The direction 
of this (1,1) attractor (one reflection at the surface and one reflection at the vertical 
sidewall) is counter-clockwise in agreement with dominant focusing effects in bucket geometry \cite{Maas1995,Maas1997}.

\begin{figure}
\includegraphics[width=8truecm]{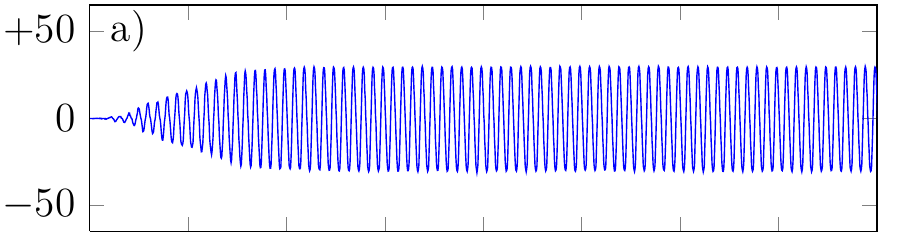}
\includegraphics[width=8truecm]{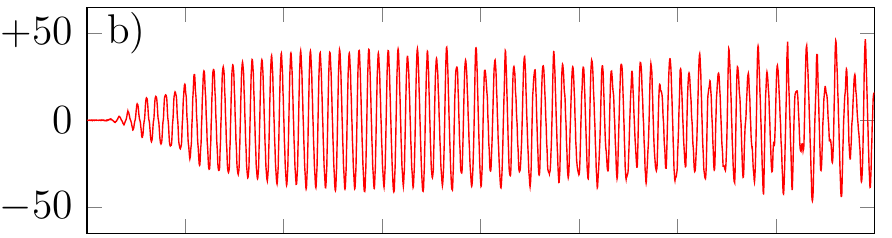}
\includegraphics[width=8truecm]{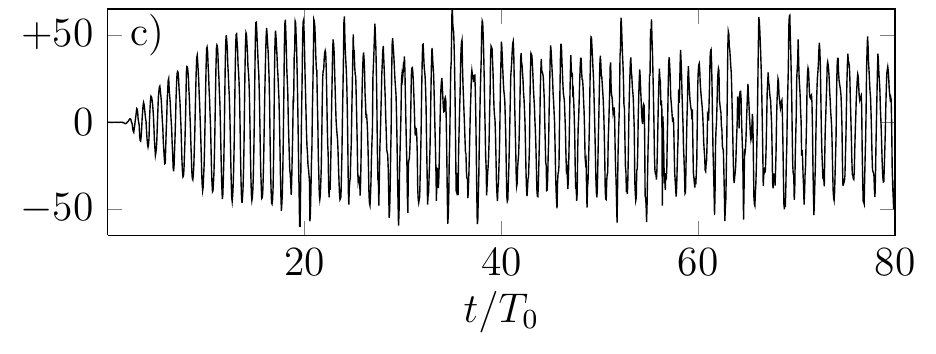}
\caption{(Color online)
 Evolution versus time of the amplitude of $\partial_x\tilde\rho$ in the focusing branch of the internal wave attractor,
measured in point A defined in Fig.~\ref{rawfield}d.
The different panels correspond to different values of the injected energy
measured through the amplitude of the cames:
$a=$0.2 cm~(a), 0.25~cm~(b) and 0.5 cm~(c).
\label{figHilbert}}
\end{figure}

At a later stage, which is emphasized by Fig.~\ref{rawfield}d presenting the snapshot $t=50 T_0$,
an instability builds up in the most energetic (focusing) branch of the wave attractor.
Figures~\ref{figHilbert}
show precisely the time-serie recorded in the focusing branch of the wave attractor.
It is clearly apparent that once the generator has been switched on, the amplitude of
the horizontal density gradient field increases: then, as one could expect from
the inspection of the first three snapshots of Figs.~\ref{rawfield}, an equilibrium value
is reached after slightly more than 20 periods.
However, it is also visible that, for amplitude $a$ larger than 0.25,  the motion is much less regular in a later stage,
corresponding presumably to a superposition of several components.

Figure~\ref{rawfield}d reveals that
the instability develops in form of oblique distortions of the wave beam, reminiscent of a typical pattern of parametric subharmonic
instability (PSI) via triadic resonance~\cite{Joubaud2012,Bourget2013}. The studies on wave-wave interactions, including triadic resonance, 
has a long history~\cite{Phillips1981}. The significance of triadic resonance among other
possible mechanisms of internal-wave instability in oceanographic applications is a debated issue~\cite{Sutherland2006}. However, there is a growing body of evidence~\cite{Joubaud2012,Bourget2013,KoudellaStaquet,JenniferKraig,Mattew} that  it is a major mechanism of instability in many practical circumstances.
Energy transfer from the primary wave to two secondary waves is known to be possible when wave frequencies and wave vectors satisfy both the temporal
\begin{equation}\omega_0=\omega_1+\omega_2\label{temporalresonance}\end{equation}
and the spatial
\begin{equation}{\bf k}_0={\bf k}_1+{\bf k}_2\label{spatialresonance}\end{equation}
conditions for triadic resonance, where subscripts 0, 1 and 2 refer to the primary, and both secondary waves, respectively.
Let us check the fulfillment of Eqs.~(\ref{temporalresonance}) and~(\ref{spatialresonance}) in our case.

\begin{figure}
\includegraphics[width=\linewidth]{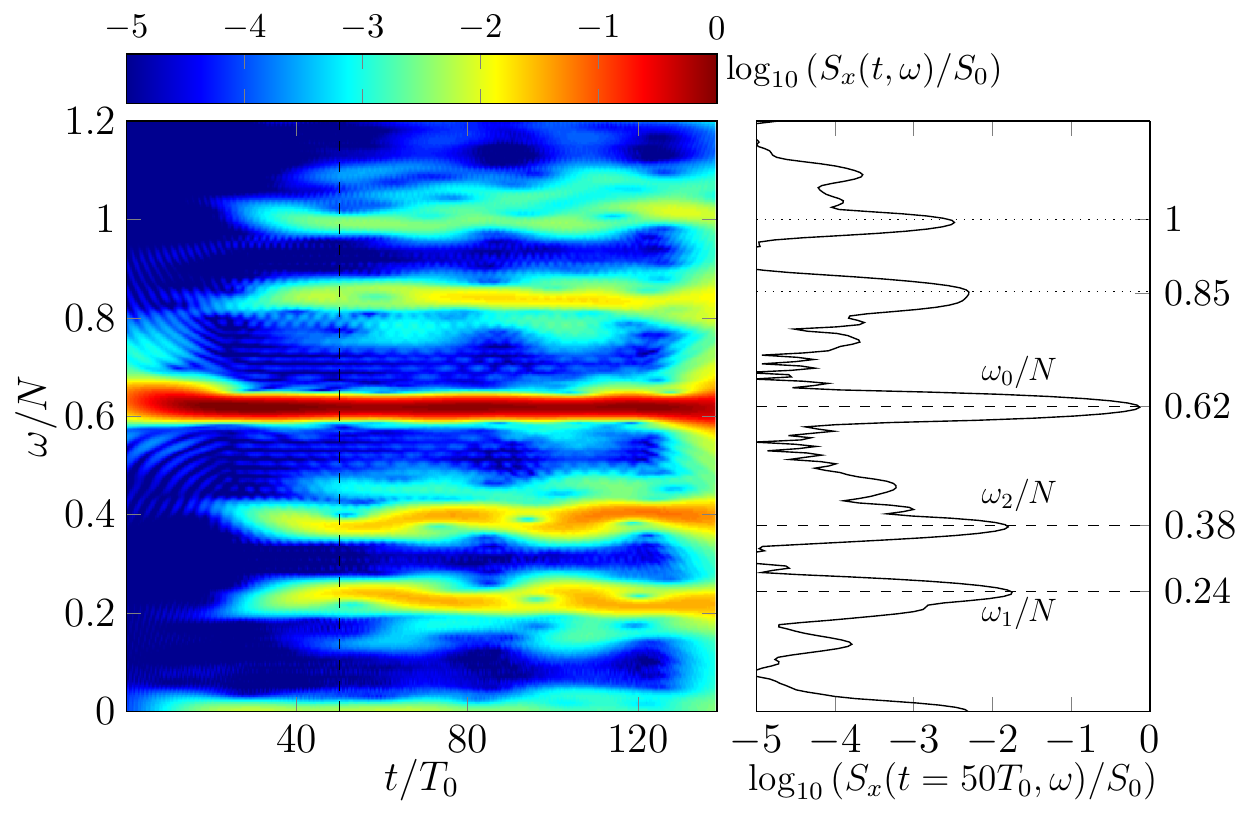}
\caption{(Color online) (a) Time frequency spectrum $S_x(\omega,t)$ of the horizontal density 
gradient field for $a=0.25$ cm.  Data were averaged on the small white rectangle shown in Fig.~\ref{rawfield}(c).
(b) Frequency spectrum
$S_x(\omega,t = 50T_0)$. The quantity $S_0$ is defined as the time average of the main component 
$ \langle S_x(\omega_0, t)\rangle$.
\label{figtimefrequency}}
\end{figure}

Figure~\ref{figtimefrequency} presents the time-frequency spectrum, defined as in \cite{Joubaud2012,Bourget2013},
of an area in the focusing branch of the attractor.
This picture confirms that the amplitude of the main frequency component reached quickly its asymptotic value, however
this representation emphasizes also that the frequency content is very rich.
The main frequency components revealed
via time-frequency analysis are listed in Table~\ref{tab:table1}. The measured 
frequency of the primary wave is equal to the forcing frequency $\omega_0$,
while the values for the secondary waves show that the temporal resonance 
condition~(\ref{temporalresonance})  is satisfied with a good accuracy. All frequencies satisfy the dispersion relation individually.
For the sake of completeness, note that two smaller peaks are also visible in Fig.~\ref{figtimefrequency} thanks to the
logarithmic scale. Corresponding to the buoyancy frequency~$N$ and $N\cos\alpha$,
 the oscillations along the inclined wall, these two natural frequencies are excited as expected for a nonlinear system.

\begin{table}[b]
\begin{ruledtabular}
\begin{tabular}{cccccc}
Location&Subscript&\textrm{$\omega/N$} & $\ell$ & $m$ & $\left|{\bf k}\right|$\\
&& & [\textrm{m}$^{-1}$]& [\textrm{m}$^{-1}$]& [\textrm{m}$^{-1}$] \\
\colrule
Injection & * &0.62 &  {+7.6}   & $\pm$ {9.6} & {12.3}\\
Attractor &0 &0.62 & {-64 ($\pm 1$)}& {-76  ($\pm 1$)} & 99\\
Attractor &1 &0.24 & {+39 ($\pm$ 5)}& {+177 ($\pm 10$)}& {181}\\
Attractor &2 &0.38 & {-108 ($\pm 3$)}& {-265 ($\pm 3$)} & 287\\
 \end{tabular}
\end{ruledtabular}
\caption{Main frequency values {of the attractor} determined with the time-frequency values and the corresponding
values of the wave vector components $(\ell, m)$ that have been measured with the Hilbert 
transform~\cite{Mercier2008}. Both components of the wave vector have been computed with 
${\bf k}=-\nabla \phi$ where $\phi$ corresponds to the phase shown in Fig.~\ref{Hilberttransform}. 
Errors in measurements of $\bf k_1$ are larger as phase lines are more horizontal and the 
measurement zone near the slope is smaller. Characteristics of the initial injection are 
calculated from the frequency value and the vertical wave number $m_*=\pi/H=9.64~\textrm{m}^{-1}$. }
\label{tab:table1}
\end{table}

\begin{figure}
\includegraphics[width=\linewidth]{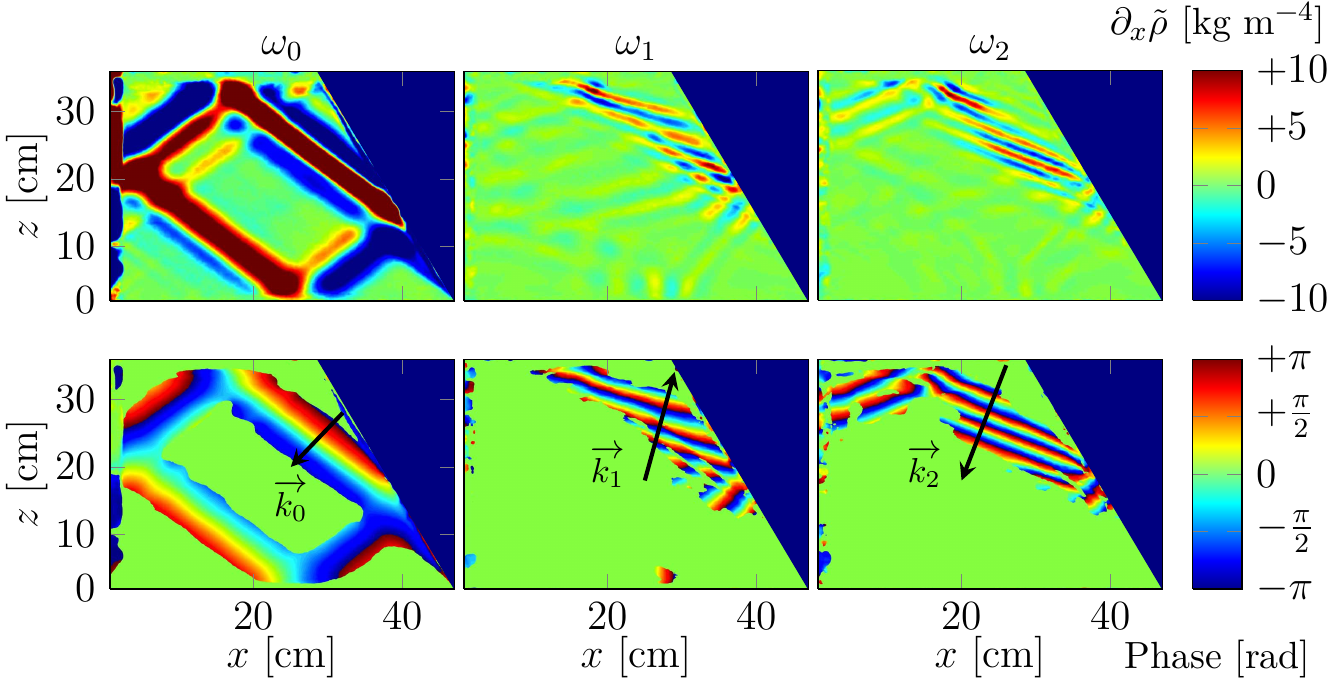}
\caption{(Color online) The top (respectively bottom) row presents the real part
(resp. phase) of the Hilbert transform at $t=50T_0$. Each column corresponds to 
a filtering around the following three frequencies:
$\omega_0$, $\omega_1$ and $\omega_2$ in the first, second and third columns 
respectively. The phase is displayed only where the wave amplitude |$\partial_x\tilde\rho$| is larger than 15$\%$ of the maximum.
\label{Hilberttransform}}
\end{figure}

The Hilbert transform, first introduced for internal-wave analysis in~\cite{Mercier2008},
is a  powerful tool for the investigation of PSI, especially
to analyze the spatial resonance condition~\cite{Joubaud2012,Bourget2013}.
The results of filtering the raw data at frequencies $\omega_0$, $\omega_1$ and $\omega_2$  are presented in Fig.~\ref{Hilberttransform}.
The corresponding numerical data on the components of the primary and secondary wave vectors can be readily obtained
by differentiating the phase with respect to both spatial variables; results are presented in Table~\ref{tab:table1}.
It can be seen that the spatial resonance condition~(\ref{spatialresonance}) is satisfied with a
reasonable accuracy {($\ell_0 \simeq \ell_1+\ell_2$ and $m_0 \simeq m_1+m_2$)}. 
In a uniformly stratified infinite wave guide of depth~$H$, the injection of energy~(\ref{profile}) 
generates a horizontally propagating first mode wave which can be represented as a sum of 
two oblique waves with opposite vertical wave numbers, which parameters are given in the first line of Table~\ref{tab:table1}.
The data of wave vectors involved in the triadic resonance show that
the combination of the wave attractor with PSI provides an extremely efficient
transfer from large to small length scales, namely from 12 to 290 m$^{-1}$ in wave numbers:
the length of the secondary waves is
roughly 25 times~(!) shorter than the scale at which the energy is injected into the system.
Note that the global Reynolds number in the
experiment is  $\displaystyle Re={a \omega_0 H}/{\nu}\simeq500$, where $\nu$ is the kinematic viscosity.
In natural systems characterized by much larger values of the Reynolds number,
we can expect an even larger difference between the length scales of input
perturbation and secondary waves, which attests the quite dramatic energy transfer at play here~\cite{Bourget2013}.

\begin{figure*}
\includegraphics[width=\linewidth]{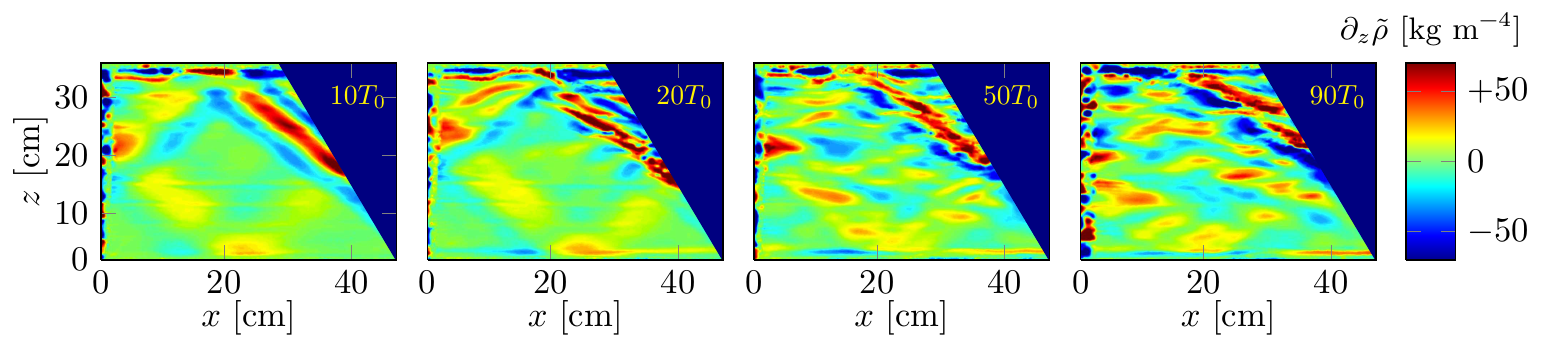}
\caption{(Color online) Snapshots of the vertical density gradient field  for $t=10T_0$~(a),  $20T_0$~(b),  $50T_0$~(c) and  $90T_0$~(c)
for a larger forcing amplitude $a=5$ mm.
Note that the shade scale (color scale online) is the same in all panels.
\label{rawfieldlargeamplitude}}
\end{figure*}

This efficient energy transfer to short length scales occurs despite
the rather small amplitude of the initial perturbation: indeed,
the non-dimensional value $a/H=0.008$
  leads already to a significant degradation of the wave attractor at large time of observation.
It is therefore important to stress that these experimental results are in
contrast with numerical results~\cite{Grisouard2008}
which successfully reproduced experiments of~\cite{Hazewinkel2008}.
However, in Ref.~\cite{Grisouard2008}, at large input perturbation (an order of magnitude
higher than the one used in the main run of simulations), 
authors have reported only weakly non-linear effects through wave components
excited at multiples of the forcing frequency, {\em i.e.} at $2\omega$ and $3\omega$. No fingerprints of PSI were mentioned. A time frequency spectrum and a Hilbert transform analysis (not already popularized~\cite{Mercier2008}), would have been necessary to unambiguously clarify this. They were not provided.

\begin{figure}
\includegraphics[width=\linewidth]{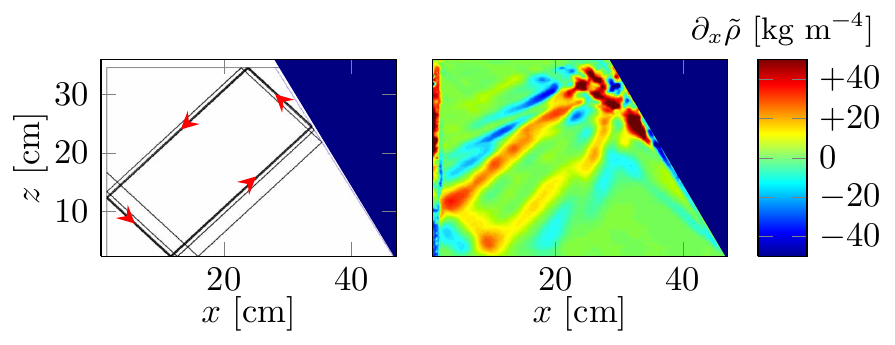}
\caption{(Color online) 
Left panel presents the prediction of a wave-ray billiard 
when $\omega_0/N=0.69 \pm 0.01$ and
$\alpha=30^{\circ}$.
Right panel  shows the snapshot of the vertical 
density gradient field for $t=30T_0$  and $a=0.25$~cm in which PSI is 
visible.
\label{diiferent_pattern}}
\end{figure}

As the amplitude of oscillation increases, the transfer of energy to short spatial scales intensifies.
Figure~\ref{rawfieldlargeamplitude} shows the evolution of the wave field at $a/H=0.015$
for a larger amplitude $a$=5 mm with the same experimental conditions and geometry depicted in Fig.~\ref{Experimentasetup}. It can be seen
that the instability sets in very quickly so that, finally,  one can hardly distinguish an attractor in the wave field which
consists of disintegrated patches and layers. For a different pattern of the attractor and high enough amplitude, PSI was also observed and the mechanism is unchanged as emphasized by Fig.~\ref{diiferent_pattern}.

Interestingly, ``patchiness'' of internal
wave beams has been reported in some oceanographic observations~\cite{vanHaren2010}.
The experimental results at the laboratory scale presented in this letter reproduce
this effect, which hinders the observation of the attractor in real oceanographic conditions.
In absence of the sloping wall, no PSI was reported for similar frequency and amplitude parameters~\cite{Bourget2013}. Consequently, the present experimental arrangement is a ``mixing box'' since it allows very efficient destabilization of the internal wave field while using relatively low amplitudes of oscillation of the generator.
 	 	
{\bf Conclusions.} Previous theoretical, numerical and experimental literature
on wave attractors is almost entirely focused on geometrical issues
and linear mechanisms. In the present letter, we consider for the first time the ultimate
instability of wave attractors. We use a new method of generation
which allows an efficient injection of energy into internal-wave attractors. Attractors are
created in a uniformly stratified fluid in a quiescent test tank with classic
trapezoidal geometry by standing-wave-type motion of a vertical boundary.

We show that the energy injected into the system by the generator nicely focused on the wave attractor. 
As the amount of energy is increased (above $a=0.2$), the attractor is destroyed by parametric subharmonic
instability (PSI) starting in the most energetic branch of the attractor and gradually eroding
its structure.
 This two-step process provides an
efficient energy transfer from the global scale associated to the size of the fluid domain
to local scales associated with the secondary waves
generated via triadic resonance. Beyond the instability, the attractor is  transformed into a
structure consisting of small-scale wave patches and layers, which hardly bear
any resemblance to the classic attractor pattern
coming from ray tracing or linear theoretical solution for the stream function.
Therefore, even in nearly perfect geometrical conditions attractors may be very hard or
impossible to observe in natural systems if the injected
energy is too large to allow the existence of a stable attractor.
Thus, the ability of attractors to concentrate wave energy places them at the
origin of a spectacular energy cascade.

\begin{acknowledgments}
EE gratefully acknowledges ENS Lyon 
for a visiting professor position 
and also partial support from grant No.~11.G34.31.0035 of Russian Government and grant 2.13.3 of RAS.
This work has been partially supported by the ONLITUR grant (ANR-2011-BS04-006-01)
and achieved thanks to the resources of PSMN from ENS de Lyon. We thank B. Bourget, S. Joubaud, P. Odier, A. Venaille for helpful discussions.

\end{acknowledgments}

\end{document}